# Exact mathematical solution for nonlinear free transverse vibrations of beams

Mohammad Asadi-Dalir[*]

Mechanical Engineering Department, Bu-Ali Sina University, 65175-4161 Hamedan, Iran

Abstract

In the present paper, an exact mathematical solution has been obtained for nonlinear free transverse vibration of beams, for the first time. The nonlinear governing partial differential equation in un-deformed coordinates system has been converted in two coupled partial differential equations in deformed coordinates system. A mathematical explanation is obtained for nonlinear mode shapes as well as natural frequencies versus geometrical and material properties of beam. It is shown that as the $s$ th mode of transverse vibration excited, the mode $2s$ th of in-plane vibration will be developed. The result of present work is compared with those obtained from Galerkin method and the observed agreement confirms the exact mathematical solution. It is shown that the governing equation is linear in the time domain. As a parameter, the amplitude to length ratio $(\Lambda/l)$ has been proposed to show when the nonlinear terms become dominant in the behavior of structure.

Keywords: Exact mathematical solution, geometrically nonlinear terms, deformed coordinates, beam, mode shape,

1. Introduction

Analysis of mechanical structures such as beams, plates and shells has been the subject of numerous researches due to their wide applications in industry. These researches are formulated based on theories and methods long evolved. To explain and prove the problems, beam is a good suggestion due to its simple equations. However the governing equations of all mechanical structures follow common principles arisen from their theorizing. The problem of stress of beam under poor bending was a challenging problem in its contemporary period. The famous mathematician L. Euler offered its formulation for the first time, based on which, J. Bernoulli [1] who was Euler's assistant, developed his analogy for the beams vibration with only bending rigidity. The error of results in Euler-Bernoulli theory was increased in higher modes and thicker structures, since no transverse shear stress was included. For this reason, this theory forecasts the natural frequency more than real value and deflection of the beam less than real value. To consider the effect of shear deformation, Timoshenko [2, 3] formulated beam equation in which transverse shear strain was constant along thickness. Shear stresses in Timoshenko theory were constant across the beam thickness and their values were equal to those of mid-plane while in real case, shear stress is a second order function of thickness, and its value at the bottom and top of its surface is zero. Therefore, this theory demanded applying a shear correction coefficient to modify. Reddy [4] proposed a third order shear deformation theory (TSDT) in which shear stress is the second order function of thickness and in top and bottom surfaces, its value is zero. So, there is no need to apply shear correction coefficient in TSDT.

[*] Corresponding author Tel.: +98 8138272410.
E-mail address: radan.dalir@yahoo.com (M. Asadi-Dalir).

When the lateral deflection of beam is large, in-plane forces are important in transverse vibration. In this case bending and stretching stiffness have interaction with each other [5]. Pai and Nayfeh [6] showed that Von-Karman strains cannot be used to derive fully nonlinear beam model. To fully account geometry nonlinearity, the second Piola-Kirchhoff stresses should be used instead of Cauchy stresses. However, in this case, yet, the evaluated stretching effect is not consistent by boundary condition. Since the stiffness is dependent upon material property in fiber reinforced beams, if stress and strain are defined with respect to un-deformed coordinates, the geometrical nonlinearity can be mistaken as material nonlinearity. Consequently, Pai and Nayfeh [7] showed that local stress and strain are required. A number of researchers focused their effort to find exact relation of deformed and un-deformed coordinates to model geometrical nonlinearities exactly [8-19]. The main problem in these researches was that the finite rotation in Euler coordinates transformation are neither independent from each other nor along a three orthogonal axis. Alkire [12] showed that different sequence of Euler rotations result in different equations of motion. Ho, Scott and Eisley [20] using Green strain in longitudinal direction, investigated large amplitude motion of simply supported beam. Heyliger and Reddy [21] and Sheinman and Adan [22], by applying Von-Karman strains, investigated large deflections of beam. Bolotin [23] and Moody [24] showed that nonlinear inertia effects are not as significant as nonlinear elasticity effects. However, Crespo da Silva and Glyn [11, 25] showed that generally ignored nonlinear terms, deduced by curvature, are of the same order of inertia nonlinearity terms and they might have a remarkable effect on behavior of structure. Nayfeh and Pai [26] showed that the nonlinear terms in mechanical structures are of hardening type and dominate in lower modes. On the other hand, they showed that the nonlinear inertia terms are of softening type and become more effective as modes increase. Many other valuable researches conducted by other scientists in the account of influence of nonlinear terms on beam behavior [27].

Other researches on the nonlinear vibration behavior of beams have been performed recently. Ahmed and Rhali [28] stablished a theoretical framework for the nonlinear transverse vibration of Euler-Bernoulli beams in which a finite number of masses were placed on arbitrary points of beam's length. Wang et.al [29] investigated principal parametric resonance of axially accelerating hyperplastic beam. Seddighi and Eipakchi [30] applied the multiple scales method to study the dynamics response of an axially moving viscoelastic beam with time dependent speed. Casaloti et.al [31] studied multi-mode vibration absorption capability of a nonlinear Euler-Bernoulli beam. Flexural vibration superposition of Euler-Bernoulli beam was investigated by [32]. Asghari et.al [33] reviewed a nonlinear size-dependent Timoshenko beam model based on the modified couple stress theory. Lewandowski and Wielentejczyk [34] studied the problem of nonlinear, steady state vibration of beams, harmonically excited by harmonic forces. The problem of geometrically nonlinear steady state vibrations of beams excited by harmonic forces investigated by [35]. Roozbehani et.al studied the nonlinear vibrations of beams by considering Von-Karman's nonlinear strains and shear deformable theory of Timoshenko. In this research, the beam is excited by applying a suddenly electrostatic force [36]. Alipour et.al showed an analytical solution for nonlinear vibration of beams having been actuated electrostatically [37]. Stojanpvoc investigated the nonlinear vibrations of Timoshenko beams on nonlinear elastic foundation by considering geometrically nonlinearities [38].

Based on the published literature, there is not any specific way to determine exact mathematical solution for geometrical nonlinear vibration of beams. In this paper, the relation between un-deformed and deformed coordinates of beam is determined, since, as it is shown; geometrical nonlinear terms are projections of axial stresses in vertical direction. The nonlinear PDE in Lagrange view is shown to be convertible in two linear PDEs in Euler view. These equations have interaction with each other and as a result have been solved together. The mode shapes as well as natural frequency of nonlinear transverse vibration of beam have been obtained for the first time. It is shown that the ordinary differential equation of beam is linear in time. The beam is considered to be a 2D one, under nonlinear elasticity and Euler-Bernoulli hypothesis, in which large deflections result in stretching its length. The results for the example of a pinned-pinned beam are compared with those obtained from Galerkin method and good agreement observed. The effect of amplitude on mode shapes and natural frequency reviewed and the results has been discussed.

2. Governing equation

2.1. Lagrange coordinates system

Consider the beam shown in Figure 1, whose geometrical and material properties are: width $b$, thickness $h$, length $l$, density $\rho$ and elasticity modulus $E$. The deformation field of beam under Euler-Bernoulli hypothesis is as below [39]:

$$u(x,z,t) = u_0(x,t) - z\frac{\partial w_0(x,t)}{\partial x}, \tag{1}$$

$$w(x,z,t) = w_0(x,t). \tag{2}$$

In Eq. (1) and (2) $u_0$ and $w_0$ are deformations of mid-plane, also $u$ and $w$ are deformation of any arbitrary point along $x$ and $z$ axis, respectively.

Now, consider the Lagrangian $x, y, z$ and Eulerian $x^*, y^*, z^*$ coordinates systems with a common origin before deformation (Figure 2). Equations of motion are derived in Lagrangian coordinates system because it has a fixed origin. In the linear case, on each face of the infinitesimal element there is only one component of stress which is in the direction of the coordinate axes. However, in the nonlinear case, there are three effective stress components which are in the direction of the coordinate axes. As shown in Figure 2, $\sigma_{xz}$ is the only stress component in the $z$ direction in the linear case whereas for the nonlinear case, $\sigma_{xx}\partial w/\partial x + \sigma_{xy}\partial w/\partial y + \sigma_{xz}$ are applied in this direction. As a consequence, it can be noted that the nonlinear terms in equations of motion are the projects of in-plane stresses which are resulted due to the large slopes. So, nonlinear governing equations in lagerangian coordinate system can be expressed as [40]:

$$\frac{\partial}{\partial x}\left(\sigma_{xx} + \sigma_{xy}\frac{\partial u}{\partial y} + \sigma_{xz}\frac{\partial u}{\partial z}\right) + \frac{\partial}{\partial y}\left(\sigma_{yx} + \sigma_{yy}\frac{\partial u}{\partial y} + \sigma_{yz}\frac{\partial u}{\partial z}\right) + \frac{\partial}{\partial z}\left(\sigma_{zx} + \sigma_{zy}\frac{\partial u}{\partial y} + \sigma_{zz}\frac{\partial u}{\partial z}\right) + f_x = \rho u_{,tt}, (3)$$

$$\frac{\partial}{\partial x}\left(\sigma_{xx}\frac{\partial v}{\partial x} + \sigma_{xy} + \sigma_{xz}\frac{\partial v}{\partial z}\right) + \frac{\partial}{\partial y}\left(\sigma_{yx}\frac{\partial v}{\partial x} + \sigma_{yy} + \sigma_{yz}\frac{\partial v}{\partial z}\right) + \frac{\partial}{\partial z}\left(\sigma_{zx}\frac{\partial v}{\partial x} + \sigma_{zy} + \sigma_{zz}\frac{\partial v}{\partial z}\right) + f_y = \rho v_{,tt}, (4)$$

$$\frac{\partial}{\partial x}\left(\sigma_{xx}\frac{\partial w}{\partial x} + \sigma_{xy}\frac{\partial w}{\partial y} + \sigma_{xz}\right) + \frac{\partial}{\partial y}\left(\sigma_{yx}\frac{\partial w}{\partial x} + \sigma_{yy}\frac{\partial w}{\partial y} + \sigma_{yz}\right) + \frac{\partial}{\partial z}\left(\sigma_{zx}\frac{\partial w}{\partial x} + \sigma_{zy}\frac{\partial w}{\partial y} + \sigma_{zz}\right) + f_z = \rho w_{,tt}. (5)$$

The above equations are well-known Lagrangian equations using Kirchhoff stress components in nonlinear elasticity and no one of Euler-Bernoulli beam assumptions has been applied, yet. Density $\rho$ is considered to be constant. To obtain Euler-Bernoulli beam equations we should consider: 1. all of stress components can be neglected compared with $\sigma_{xx}$ and $\sigma_{xz}$, 2. the existing variables in governing equations which are resulted from beam deformation field (Eq. (1) and (2)) are only unknown functions of $x$ and are determined with respect beam area, and 3. the initially perpendicular straight lines to mid-plane remain straight and perpendicular after deformation, as well. In addition to the mentioned assumptions, in studying the transverse vibrations of mechanical structures, the terms $\partial u/\partial x_i$ and $\partial v/\partial x_i$ are negligible compared with $\partial w/\partial x_i$. As is observed in Figure 3-b, the transverse deflection of the beam only can induce the nonlinear terms involving $\partial w / \partial x$ and the other two slopes ($\partial u / \partial x_i$ and $\partial v / \partial x_i$) become remarkable when the beam undergoes large in-plane deformations.

Using the third assumption we have:

$$Q_x = \frac{\partial M_x}{\partial x}. \tag{6}$$

The variables in Eq. (6) are defined as $Q_x = \int \sigma_{xz} dA$ and $M_x = \int z.\sigma_{xx} dA$ (These stress components can be seen in the first term of Eq. (5)). In the governing equations (3-5) the differentials are with respect to $x$, $y$ and $z$, while the variables are only unknown functions of $x$. Therefore, by integrating on the beam area, the governing equations would be only in terms of differentials with respect to $x$. Applying beam assumptions, neglecting body forces $f_i$ and replacing Eq. (6) into (5) we achieve the governing equations in the following form:

$$\frac{\partial N_x}{\partial x} = m \frac{\partial^2 u_0}{\partial t^2}, \tag{7}$$

$$\frac{\partial}{\partial x}\left(\frac{\partial M_x}{\partial x} + N_x \frac{\partial w_0}{\partial x}\right) = m \frac{\partial^2 w_0}{\partial t^2}. \tag{8}$$

One can find that Eq. (2) is used in Eq. (8) (i.e. $\partial w / \partial x = \partial w_0 / \partial x$). In these equations the new variables are $N_x = \int \sigma_{xx} dA$ and $m = \int \rho dA$. To resolve the governing equations exactly, we should have a better understanding of the existing terms in Eq. (7) and (8). Each term in above equations has been appeared due to a particular type deformation in the structure. The strain field in nonlinear vibration is $\varepsilon_x = \partial u / \partial x + (\partial w / \partial x)^2 / 2$ [39], where $u$ is defined by Eq. (1). In Eq. (7), $N_x$ arisen from $u_0$ and play no role in transverse vibrations. Therefore Eq. (7) can be neglected when $u_0 = 0$. However, $M_x$ in Eq. (8) arisen from bending effect and deduced by the second part of Eq. (1), play the main role in the transverse vibration. The most important term in the nonlinear transverse vibration is $N_x \partial w_0 / \partial x$. Although it is deduced by an in-plane deformation, it just appears in transvers vibration and plays no role in the in-plane vibration. The geometrical reason of this term depicted in the Figure 3. In Figure 3-A, as can be observed, the large deflection lead to increasing length of beam and the resulted strain can be considered as: $(dS - dx)/dx = (\partial w / \partial x)^2 / 2$ which was shown in strain field, already. As a result, although $u_0 = 0$, the stress component $\sigma_{xx}$ is developed. Since $\partial w / \partial x$ is big, regarding Figure 3-B, in this case $\sigma_{xx}$ is effective in transverse vibrations.

2.2. Euler coordinates system

Consider Figure 2 in which the components of stress field constitute linear governing equations in the $x^*, y^*, z^*$ coordinates system. It is obvious that the governing equations in this coordinates system are very simple. In three dimensional theory of elasticity the relation between two coordinates system is unknown. However here, as shown in Figure 3-B, the normal stress is along beam length appearing in $z$ direction due to large deflection. It is clear in Figure 4 that the rotation value of the Euler configuration with respect to Lagrange configuration is equal to $\partial w / \partial x$. As a consequence, the problem is linear in $x^* - z^*$ coordinates system. This coordinates system has been located on mid-plane. The deformation field variables in Eulerian coordinates system are $u_0^*$ and $w_0^*$ the deformations of mid-plane, and also $u^*$ and $w^*$ the deformations of any arbitrary point of beam along $x^*$ and $z^*$ directions, respectively. Although Eq. (7) can be neglected when $u_0 = 0$, this equation should be included in Eulerian coordinates system yet, since $u_0^* \neq 0$. Because the term deduced by increasing length of beam $N_{x^*}$, in Eulerian coordinates system appear in $x^*$ direction and not $z^*$. The governing equations in Elerian coordinates system are in the following form:

$$\frac{\partial N_{x^*}}{\partial x^*} = m\frac{\partial^2 u^*_0}{\partial t^2}, \tag{9}$$

$$\frac{\partial^2 M_{x^*}}{\partial x^{*2}} = m\frac{\partial^2 w^*_0}{\partial t^2}. \tag{10}$$

As a proof, these equations can be obtained simply using a well-known infinitesimal element in which regarding Figure 4, transverse shear stress is along $z^*$ and normal stress is along $x^*$.

3. Exact mathematical solution

The governing equation of nonlinear transvers vibration of beam in Lagrange coordinates system can be obtained by replacing the corresponding form of Eq. (8) in terms of strain field components [39] in following form:

$$-EI\frac{\partial^4 w_0}{\partial x^4} + \frac{3}{2}AE\left(\frac{\partial^2 w_0}{\partial x^2}\right)\left(\frac{\partial w_0}{\partial x}\right)^2 = m\frac{\partial^2 w_0}{\partial t^2}. \tag{11}$$

The above equation is a nonlinear partial differential equation and has no exact mathematical solution due to presence of nonlinear term. In this equation, $I$ and $A$ are surface inertia and area of beam, respectively. We remind Figure 3, the applying forces in nonlinear vibration are shown in Figure 5. In this figure, the vector $R$ can be representative of left hand side of Eq. (8). It is obvious that Eq. (11) can be decomposed in two coupled equations in below form:

$$EA\frac{\partial^2 u^*_0}{\partial x^{*2}} = m\frac{\partial^2 u^*_0}{\partial t^2}, \tag{12}$$

$$-EI\frac{\partial^4 w^*_0}{\partial x^{*4}} = m\frac{\partial^2 w^*_0}{\partial t^2}. \tag{13}$$

In the above equations, $u^*_0$ is the stretch value arisen from changing length and in-plane deformation is neglected $(u_0 = 0)$. The boundary conditions for simply supported beam cause the deformation in Eq. (12) and also moment and deflection in Eq. (13) to be equal to zero:

$$u^*(x^*,t) = w^*(x^*,t) = \frac{\partial^2 w^*(x^*,t)}{\partial x^{*2}} = 0, \qquad x^* = 0, l^*. \tag{14}$$

In the Eq. (14) we have: $l^* = \int_0^l \sqrt{1+(\partial w/\partial x)^2}\, dx$. For the linear transverse vibration, Eq. (13) is resolved independently and when the in-plane vibration is considered, Eq. (12) should be resolved. However, for the present problem, both equations are effective simultaneous and have interaction with each other. Therefore, in the nonlinear free vibrations, in addition to satisfying Eq. (12-14), the geometry of deformation seen in Figure 5 should be satisfied, as well. By considering the method of separation of variables, to separate them, we consider the following functions:

$$u^*_0(x^*,t) = U^*(x^*)\sin(\omega t), \tag{15}$$
$$w^*_0(x^*,t) = W^*(x^*)\sin(\omega t), \tag{16}$$
$$w(x,t) = W(x)\sin(\omega t). \tag{17}$$

The three variables $u^*_0(x^*,t)$, $w^*_0(x^*,t)$ and $w(x,t)$ constitute a deformation field simultaneous. As a result, the time dependent function is considered for them $\sin(\omega t)$. Meanwhile, the shown geometrical constraint in Figure 5 $(\vec{W^*} + \vec{U_i} = \vec{W})$, induce the following equations:

$$U_i = \left| W^* \frac{dW^*}{dx^*} \right|, \qquad \left( \frac{\partial w^*}{\partial x^*} \cong \frac{\partial w}{\partial x} \right), \tag{18}$$

$$W(x) = \sqrt{(U_i(x^*))^2 + (W^*(x^*))^2}. \tag{19}$$

In Eq. (18) the variable $U_i$ is considered to be only a positive number. Because, for the immovable boundary conditions that have been considered here, the beam only is allowed to be stretched. Since the Eq. (12) and (13) are to be solved simultaneous, It may the mode shapes of $U^*$ cannot satisfy the geometry constraint of Figure 5. Hence, the considered $U_i$ in Eq. (18) achieved from combination of several modes of $U^*$. In fact, the shown constraint of Figure 5 determines $U^*$ so that, the resultant vector in Eq. (19) is vertical. Replacing Eq. (15) and (16) into Eq. (12) and (13) respectively, one can find separately the solution of them for boundary conditions in Eq. (14) in the following form [41]:

$$U_r^*(x^*) = A_r \sin\left(\frac{r\pi}{l^*}x^*\right), \qquad \omega_r^2 = \frac{EA}{m}\left(\frac{r\pi}{l^*}\right)^2, \qquad (20)$$

$$W_s^*(x^*) = A_s \sin\left(\frac{s\pi}{l^*}x^*\right), \qquad \omega_s^2 = \frac{EI}{m}\left(\frac{s\pi}{l^*}\right)^4. \qquad (21)$$

The independent solution obtained, while the interaction between them should be included. Inserting Eq. (21) into (18), we find:

$$U_i = \left| A_s^2 \frac{s\pi}{2l^*} \sin\left(\frac{2s\pi}{l^*}x^*\right) \right|. \qquad (22)$$

We realize from Eq. (22) that as the $s$th mode of transverse vibration excited $(W_s^*)$, the mode $r = 2s$ th from in-plane vibration $(U_r^*)$ appears simultaneous. The combination of them gives $s$th mode of nonlinear transverse vibrations $(W(x))$. Therefore, in the nonlinear transverse vibration of beam for $r = 2s$, the governing equation and boundary conditions are satisfied as well as geometrical constraint of Figure 5. If we consider the weight coefficient of contribution of in-plane modes $U_r^*$ in the nonlinear transverse vibration as $c_r$ $\left(U_i = \sum_{r=1}^{\infty} c_r U_r^*\right)$, Eq. (22) implies that:

$$c_r = 1 \ (r = 2s), \quad \text{and} \quad c_r = 0 \ (r \neq 2s). \qquad (23)$$

Using Eq. (21) and (22) in (15-17) results in the following equations:

$$u^*_0(x^*,t) = \left| \Lambda^2 A_s^2 \frac{s\pi}{2l^*} \sin\left(\frac{2s\pi}{l^*} x^*\right) \sin(\omega t) \right|, \qquad (24)$$

$$w^*_0(x^*,t) = \Lambda A_s \sin\left(\frac{s\pi}{l^*} x^*\right) \sin(\omega t), \qquad (25)$$

$$w(x,t) = \Lambda A_s \sin\left(\frac{s\pi}{l} x\right) \sqrt{1 + \left(\Lambda A_s \frac{s\pi}{l} \cos\left(\frac{s\pi}{l} x\right)\right)^2} \sin(\omega t). \qquad (26)$$

In the Eq. (24-26), the variable $\Lambda$, is inserted to consider the effect of amplitude in nonlinear transverse vibration. In fact the resultant of $\Lambda A_s$, as the maximum deflection of beam along $z^*$, represents the amplitude, because the normalized mode shapes are used. It is observed that, although Eq. (24) and (25) are versus $x^*$, Eq. (26) is a function of $x$ because $w$ is perpendicular to $x$. Eq. (20) and (22) represent that the beam vibrates with natural frequency of $\omega_r$ as a rod and its natural frequency will be $\omega_r(r=2s)$ if the mode shape arisen from Eq. (18) is imposed to the structure, due to beam's stretching stiffness. On the one hand, according to Eq. (21) the beam vibrates with the natural frequency of $\omega_s$ due to its bending stiffness in transverse vibration. However in the nonlinear free vibrations two equations are coupled to each other and both stretching and bending stiffness appear in natural frequency. Let us assume the stretching stiffness of beam is $k_r$ in a way that $\omega_r^2 = k_r/m$, and bending stiffness is $k_s$ so that $\omega_s^2 = k_s/m$. Here author emphasizes that $k_r$ and $k_s$ are both linear stiffness. A schematic model of system represented in Figure 6 for more clarity of concept. It can be concluded from physics of problem, similar to Figure 6, that the springs are parallel to each other but their amplitudes are unequal. Noticing to Figure 6 we find that the resultant stiffness of nonlinear vibrations is $k = k_s + \lambda^2 k_r$ and the resulted natural frequency will be $\omega^2 = \omega_s^2 + \lambda^2 \omega_r^2$. In the recent equation, $\lambda$ is the ratio of amplitude of in-plane vibration with respect to that of transverse vibrations (i.e. $\Lambda A_s$). To find that, one can compare Eq. (24) and (25) which shows: $\lambda = A_s \Lambda \left( s\pi / 2l^* \right)$. Thus, the nonlinear natural frequency of beam $\omega$ can be obtained in following shape:

$$\omega = \left(\frac{s\pi}{l^*}\right)^2 \sqrt{\frac{E}{\rho}\left((A_s \Lambda)^2 + \frac{h^2}{12}\right)}, \qquad (s = 1, 2, 3, \ldots). \tag{27}$$

As will be discussed soon, the above natural frequency belongs to a hardening nonlinear system. For the present problem nonlinear inertia has been neglected because it is not excited. However, for movable boundary conditions, such as a clamped-free beam, in-plane inertia become effective in transverse vibration and the problem in this case will be of softening nonlinear type.

4. Results and discussion

We notice that Eq. (26) does not satisfy the Eq. (11). The reason is found in the simplifications performed during extraction process of governing equations of Lagrange coordinates, including one considered in Von-Karman strain field:

$$\frac{(dS - dx)}{dx} = \frac{\sqrt{1 + \left(\frac{\partial w}{\partial x}\right)^2} dx - dx}{dx} \cong \frac{1}{2}\left(\frac{\partial w}{\partial x}\right)^2. \tag{28}$$

The approximation used in Eq. (28) is that only two terms in Taylor expansion included. This restriction was appeared since we investigated the problem in Lagrange coordinates system. Hence, even the exact solution of Eq. (11) will have less accuracy compared with Eq. (26).

4.1 Mode shape

According to the literature of research and to the best knowledge of author, the governing equation in Lagrange coordinates system has had no exact mathematical explanation for mode shape of beam. In the approximate methods such as Galerkin and Rayleigh-Ritz methods, the linear mode shapes, satisfying boundary conditions, can be used [41]. Therefore, to verify Eq. (26) we consider a fundamental origin in scientific theorizing. Consider two theories, the one is general and covers an extensive spectrum of problems, while the other is restricted to a number of particular problems. Thus the general theory covers the problems of the other theory. In this case, the limit of the results obtained from general theory tends to that obtained from other theory when the particular problems are reviewed. As a consequence, the mode shapes shown in Eq. (26) should be the same as those of linear problem when $\Lambda \to 0$. This is shown in Figure 7. It is seen that as the $\Lambda$ decreases, the mode shape tends to the shapes of that for linear vibration. The effect of nonlinear terms in several modes of vibration is depicted in Figure 8. We notice in this figure that the nonlinear terms become evident in higher modes more than lower modes.

4.2 Natural frequency

According to the literature of problem [27], the ordinary differential equation of system is nonlinear in time domain. For a nonlinear ODE, natural frequency is a function of time $(\omega = \omega(t))$ [42]. However, Eq. (27) is constant in time domain and only varies as the amplitude of vibration, changes by $\Lambda$. In addition, Eq. (12) and (13) are linear functions of time. In fact, the nonlinear effects in Euler coordinates system were included in governing equations by applying Eq. (18). According to Eq. (18) a nonlinear mode shape is imposed to in-plane vibration of beam and based on fundamentals of vibration [41] this cannot make nonlinear the problem of longitudinal vibration. To find the reason of paradox, we review the effect of nonlinear terms on the behavior of structure and natural frequency, qualitatively.

4.2.1 The effect of natural frequency on linear resonance

Consider a one degree of freedom (DOF) system with the natural frequency of $\omega_n = 1/\pi$. In this case, the direction of the motion is changed each second. Because the natural frequency is constant and the period of the system is only a function of natural frequency, the period is constant, too. The work done by excitation force can be obtained by $\Phi = \int_0^t F \dot{w} \, dt$. If the excitation force and the motion are in the same directions, the acceleration of the system is positive; otherwise the acceleration will be negative. Consider the case in which a harmonic force $F = F_0 g(\Omega,t)$, is applied on the system. The function $g(\Omega,t)$ can be any arbitrary harmonic function. Excitation frequency $(\Omega)$ shows when the directions in the applied force changes. So, if $\Omega = \omega_n$, the applied force and the motion of the system have the same directions and energy of the system will be increased continuously $(t \to \infty \Rightarrow \Phi \to \infty)$. This phenomenon is known as resonance. In the nonlinear resonance, the same things happen except that the natural frequency changes during the motion of the beam which will be discussed in the next section.

4.2.2 The effect of natural frequency on nonlinear resonance

Linear natural frequency is only dependent on the material properties of the system. As a general definition, natural frequency of a mechanical system is the ratio of the internal driving forces to the internal inertia forces of that system. For example in a 1-DOF mass-spring system, $\omega_n = \sqrt{K/M}$ in which spring stiffness $(K)$ represents the driving forces and mass $(M)$ denotes the inertia forces which resists against acceleration.

Natural frequency of linear systems is constant which happens because the driving and inertia forces do not vary during vibration. However, in the nonlinear vibration, driving forces change with a change in the amplitude of motion. For example consider a nonlinear vibrating system with the following equation of motion: $M\ddot{f} + Kf + \alpha f^3 = 0 \left(\text{i.e. } k_1 = K, k_2 = \alpha f^2\right)$. Since the driving forces are dependent on the displacement, the effect of $\alpha$ in this equation is to change the driving forces. If $\alpha$ is positive, increasing amplitude of motion increases the driving forces and if $\alpha$ is negative, large amplitude decreases these forces which is why in the hardening nonlinear systems natural frequency increases by amplitude of the motion, whereas, natural frequency of softening systems has an inverse relation with amplitude of the motion. Nayfeh [42] expressed that the initial condition is an effective parameter on the nonlinear frequency. The proposed amplitude in this study can include initial conditions, too. The reason is that the sum of initial kinetic and potential energies determines the amplitude from maximum potential energy of the system at initial conditions. Moreover, the behavior of nonlinear frequency in terms of amplitude of vibration can be described in a better way which will be shown soon.

In free vibration, the energy of the system is constant and equal to the initial energy. However, in the forced vibration, the energy of the system changes over time. So, in the nonlinear forced vibration, the initial conditions give no information about the nonlinear frequency except at the initial time (i.e., $t=0$). But if the amplitude of motion is considered, it will be possible to analyze the nonlinear frequency at any moment of the motion.

Now, the resonance phenomenon of nonlinear systems can be explained simply. In the linear resonance (see section 4.2.1), the excitation force and the motion of system are in the same directions because the excitation frequency is equal to the natural frequency. Regarding the fact that the natural frequency of nonlinear oscillation is dependent on the amplitude of motion, resonance phenomenon increases the amplitude of motion and consequently the natural frequency will be changed. So, the system is either softening $(\omega_n < \Omega)$ or hardening $(\omega_n > \Omega)$. Thus, resonance phenomenon will be finished $(\Omega \neq \omega_n)$. This is why in the nonlinear case the excitation frequency is considered as $\Omega = \omega_n + \epsilon\sigma$ [42]. Using this scheme, the detuning parameter can vary in a way that $\Omega = \omega_{\text{nonlinear}}$ is satisfied for the whole process. This is shown in Figure 9. At the origin $(\sigma = 0)$, excitation frequency and linear natural frequency are equal to each other resulting in the resonance. In a hardening system, increase of the amplitude due to the resonance results in $\omega_n > \Omega$. So, to remain in the resonance region, $\sigma$ should be increased according to the variation of $\omega_n$. By reoccurrence of the resonance, the amplitude and subsequently the natural frequency are increased again. So, $\sigma$ should be increased again and this process is repeated frequently. Thus, for hardening case (Figure 9-b), the curve bends away from the vertical axis to the right side. On the other hand, in the softening nonlinear systems, increasing amplitude leads to $\omega_n < \Omega$. So, in these systems $\sigma$ should be decreased with the change of $\omega_n$ and as a result, frequency response curve will be bend to the left hand (Figure 9-c).

4.2.3 Nonlinear frequency of beam

As a well-known phenomenon, the reason of which was described in pervious section, natural frequency of a nonlinear system increases as the amplitude enhances (hardening). For a nonlinear PDE (beam equation), either the stiffness is a function of time or it is independent of time. In the first case, the governing ODE is nonlinear ($k_2 = \alpha f^2$) and natural frequency changes, although the amplitude is constant $(\omega = \omega(t))$. However, for the second case, natural frequency is constant when amplitude is constant. Therefore, according to the literature, the ODE for nonlinear vibration of beam is nonlinear which result in $\omega = \omega(t)$ and according to present exact mathematical solution, the governing ODE is linear. Now, we review the problem to find the fact, first from Euler's point of view and then from Lagrange's point of view.

From Euler's point of view, regarding Eq. (27) the stiffness of beam is obtained from $k = k_s + \lambda^2 k_r$. Both bending and stretching stiffness are linear (i.e. $k_s \neq k_s(w, \Lambda), k_r \neq k_r(u, \Lambda)$). The other parameter is a function of amplitude $\lambda = \lambda(\Lambda)$, and therefore $\omega = \omega(\Lambda)$. This is arisen from a nonlinear PDE being linear over time but nonlinear in space.

From Lagrange's point of view, applying approximate methods such as Galerkin on Eq. (11) gives the same ODE discussed in section 4.2.2, $M\ddot{f} + Kf + \alpha f^3 = 0$. This equation is a nonlinear ODE whose nonlinearity order is three. The number three is appeared in this equation from $N_x \partial w_0 / \partial x$ of Eq. (8). The stress component here is $N_x$ arisen from stretching effect $\left(E(\partial w_0 / \partial x)^2 / 2\right)$. By multiplying it in slope, the resulted equation is a third order function of slope and as a result, a third order function of $f(t)$.

Let us investigate $N_x$ which is a second order function of $f(t)$ since in the Taylor expansion of Eq. (28) two terms were included. Therefore, by considering more terms in Taylor expansion, the order of resulted ODE will be for example five or seven. As we know, the governing equation of any arbitrary system represents its inherent properties and it is impossible the inherent property of a system is dependent upon the number of terms in Taylor expansion considered by researcher. On the other hand the coefficient $\partial w_0 / \partial x$ has been multiplied by $N_x$ in Eq. (8) to show its project in vertical direction, while in Euler coordinates $N_{x^*}$ has not been multiplied by slope, since it is along $x^*$. As a result, if we consider two terms in Taylor expansion, the resulted ODE will be of the second order nonlinearity in Euler coordinates, while it is of the third order nonlinearity in Lagrange coordinates system. Based on primary fundamentals in deriving equations, the governing equations should be independent of the considered coordinates system, based on which they are extracted. However, we observed the problem in the Lagrange coordinates does not satisfy this basic concept and depends on the number of terms in Taylor expansion.

To overcome the problem, here we suggest that to derive nonlinear strain field, first the time dependent function is separated:

$$\frac{d(S-x)}{dx} = \frac{\sqrt{1+\left(\frac{dW}{dx}\right)^2}dx - dx}{dx} \cong \frac{1}{2}\left(\frac{dW}{dx}\right)^2. \tag{29}$$

The above equation is, in fact, another form of Eq. (18) which is independent of time. We conclude that Eq. (28) violates the shown geometrical constraint in Figure 5. Now, considering more terms in Taylor expansion has no effect on inherent properties of system. Using Eq. (29) in governing equation and separating the time dependent function, one can obtain $\left((\partial w/\partial x)^3 = (dW/dx)^3 . f(t)\right)$:

$$-\frac{EI}{m}\frac{d^4W}{dx^4} + \frac{3}{2}\frac{AE}{m}\left(\frac{d^2W}{dx^2}\right)\left(\frac{dW}{dx}\right)^2 + \omega^2 W = 0. \tag{30}$$

The first term in Eq. (30) involves bending stiffness $k_s$ and the second term involves $\lambda^2 k_r$ stretching stiffness, the sum of which gives total stiffness $k$. Therefore, for verification of the results of present work, one can apply Galerkin method on Eq. (30) and calculate the bending and stretching stiffness. The comparison is shown in Table 1. The considered geometrical and material properties in this table are: $\left(E = 200 Gpa, h = 0.05m, b = 0.02m, l = 2m, \rho = 2700 kg/m^3, \Lambda = 0.1m\right)$. It is clear that there is a good agreement between the results. The Von-Karman strain field neither satisfies Eq. (12) and boundary condition of Eq. (14) nor the geometrical constraint in Eq. (18). For this reason it has a certain error in Table 1.

The effect of amplitude $\Lambda$ on $\omega_s$, $\lambda \omega_r$ and $\omega$ is depicted in Figure 10. As it is expected, bending stiffness is not affected by amplitude. For small deflection the bending effect is dominant but as can be observed, when the amplitude increases, the stretching effects become dominant. The effect of nonlinearity on higher modes is shown in Figure 11. It is obviously observed that the stretching effect on higher modes is more effective. The material properties of beam in these figures are the same used for Table 1.

Based on the literature of the research, it is considered that when the ratio of amplitude to thickness is $\Lambda/h > 0.5$, the nonlinear terms should be considered in governing equations [43]. However, in Eq. (27) the thickness of structure is not seen in stretching stiffness. According to the same equation, the nonlinear terms are proportional to $\Lambda/l$. The same conclusion arisen from Eq. (26). Thus the thickness appears to have no direct effect on stretching stiffness. This is shown in Figure 12. It is obvious that for a specified thickness and amplitude, the effect of nonlinear terms become more sensible as the length of beam reduces. Despite this, the effect of thickness is not necessary to be neutral. Since the bending stiffness is proportional to the second order of thickness, when the beam is thin it is very effective with respect to amplitude enhancement. This is reduced as the beam become thicker. This issue is represented in Figure 13.

5. Conclusion

The nonlinear partial differential equation of beam in Lagrange coordinates system, as the vector of resultant forces, was decomposed into its constituting vectors along beam and perpendicular to that. The result was two linear partial differential equations which were solved together since they have interaction with each other. Mode shapes and natural frequency of nonlinear vibration of beam achieved for the first time. Nonlinear mode shapes were shown to tend to that of linear vibration for small amplitudes. The natural frequency, arisen from bending and stretching stiffness was compared with those given by Galerkin method and good agreement observed. It was shown that for $s$ th mode of nonlinear vibration, $s$ th mode of bending stiffness and $2s$ th mode of in-plane stiffness contribute simultaneously. The governing ODE of system was shown to be linear in the time domain. It is shown that the effect of nonlinear terms in governing equation is proportional to amplitude to length ratio $(\Lambda/l)$.

References:

1. Bernoulli, J. "Essai théorique sur les vibrations de plaques élastiques rectangulaires et Libres", *Nova Acta Acad. Petropolit*. **5**, pp. 197–219 (1789).

2. Timoshenko, S.P. "On the correction for the shear of the differential equation for transverse vibration of prismatic bars", *Phil. Mag*., **41**, pp. 744–746 (1921).

3. Timoshenko, S.P. "On the transverse vibration of bars of uniform cross sections", *Phil. Mag.*, **43**(6), pp. 125–131 (1922).

4. Reddy, J.N. "A simple higher order theory for laminated composite plates", *J. App. Mech*., **51**, pp. 745–752 (1984).

5. Pai, P.F. and Nayfeh, A.H. "Nonlinear nonplanar oscillations of cantilever beam under lateral base excitation", *Int. J. Non. Linear. Mech*., **25**, pp. 455–474 (1990).

6. Pai, P.F. and Nayfeh, A.H. "A nonlinear composite beam theory", *Nonlinear. Dyn*., **3**, pp. 431–463 (1992).

7. Pai, P.F., Palazotto, A.N., and Greer, J.M., "Polar decomposition and appropriate strain and stresses for nonlinear structural analysis", Comp. Struct. **66**, pp. 823–840 (1998).

8. Hodges, D.H., Dowell, E.H., "Nonlinear equation of motion for the elastic bending and torsion of twisted non-uniform rotor blades", *NASA TN D-7818* (1974).

9. Hodges, D.H., "Nonlinear equations of motion or cantilever rotor blades in hover with pitch link flexibility, twist, precone, droop, sweep, torque offset and blade root offset", *NASA TM X-73* (1976).

10. Dowel, E.H., Traybar, J. and Hodges, D.H. "An experimental-theoretical correlation study of nonlinear bending and torsion deformations of a cantilever beam", *J. Sound. Vib.*, **50**, pp. 533–544 (1977).

11. Crespo da Silva M.R.M. and Glynn, C.C. "Nonlinear flexural-flexural-torsional dynamics of in-extensional beams-I. Equations of motion", *J. Struct. Mech.*, **6**, pp. 437–448 (1978).

12. Alkire, K., "An analysis of rotor blade twist variables associated with different Euler sequences and pretwist treatments", *NASA TM-84394* (1984).

13. Rosen, A. and Rand, O. "Numerical model of the nonlinear model of curved rods", *Comp. Struct.*, **22**, pp. 785–799 (1986).

14. Bauchau, O.A. and Hong, C.H. "Large displacement analysis of naturally curved and twisted composite beams", *AIAA. J.*, **25**, pp. 1469–1475 (1987).

15. Rosen, A., Loeway, R.G. and Mathew, M.B. "Nonlinear analysis of pretwisted roads using principle curvature transformation. Part I: Theoretical derivation", *AIAA. J.*, **25**, pp. 470–478 (1987).

16. Minguet, P. and Dugundji, J. "Experiments and analysis for composite blades under large deflection, Part I. Static behavior", *AIAA J.*, **28**, pp. 1573–1579 (1990).

17. Pai, P.F. and Nayfeh, A.H. "A fully nonlinear theory of curved and twisted composite rotor blades accounting for warping and tree-dimensional stress effects", *Int. J. Solids. Struct.*, **31**, pp. 1309–1340 (1994).

18. Banan, M.R., Karami, G. and Farshad, M. "Nonlinear theory of elastic spatial rods", *Int. J. Solids. Struct.*, **27**, pp. 713–724 (1991).

19. Simo, J.C. and Vu-Quoc, L. "A geometrical exact rod model incorporating shear and torsion −warping deformation", *Int. J. Solids. Struct.*, **27**, pp. 371–393 (1991).

20. Ho, C.H., Scott, R.A. and Eisley, J.G. "Nonplanar, nonlinear oscillations of a beam-I. Forced motion", *Int. J. Non. Linear. Mech.*, **10**, pp. 113–127 (1975).

21. Heyliger, P.R. and Reddy, J.N. "A higher order beam finite element for bending and vibration problems", *J. Sound. Vib.*, **126**, pp. 309–326 (1988).

22. Sheinman, I. and Adan, M. "The effect of shear deformation on post-bockling behavior of laminated beams", *J. App. Mech.*, **54**, pp. 558–562 (1987).

23. Bolotin, V.V., *The Dynamic Stability of Elastic Systems*, Holden-Day, San Francisco, California. 24 (1964).

24. Moody, P. "The parametric response of imperfect column, in Developments in Mechanics", *Proceeding of the 10-th Midwestern Mechanics Conference*, pp. 329–346 (1967).

25. Crespo da Silva, M.R.M. and Glynn, C.C. "Nonlinear flexural-flexural-torsional dynamics of in-extensional beams-II. Forced motion", *J. Struct. Mech.*, **6**, pp. 437–448 (1978).

26. Nayfeh, A.H. and Pai, P.F. "Nonlinear nonplanar parametric responses of in-extensional beam", *Int. J. Non. Linear. Mech.*, **24**, pp. 139–158 (1989).

27. Nayfeh, A.H. and Pai, P.F. *Linear and nonlinear structural mechanic*, John Wiley & Sons, Inc. (2004).

28. Ahmed, A. and Rhali, B. "Geometrically nonlinear transverse vibrations of Bernoulli-Euler beams carrying a finite number of masses and taking into account their rotatory inertia", **6**, pp. 489-494 (2017).

29. Wang, Y., Ding, H. and Chen, L. "Nonlinear vibration of axially accelerating hyper-elastic beams" *Int. J. Non. Linear. Mech.*, In Press (2018).

30. Seddighi, H. and Eipakchi, H.R. "Dynamic response of an axially moving viscoelastic Timoshenko beam", *J. Solid Mech.*, **8**, pp. 78–92 (2016).

31. Casalotti, A., El-Borgi, S. and Lacarbonara, W. "Metamaterial beam with embedded nonlinear vibration absorbers", *Int. J. Non. Linear. Mech.*, **98**, pp. 32–42 (2018).

32. Wang, T., Sheng, M. and Qin, Q. "Multi-flexural band gaps in an Euler–Bernoulli beam with lateral local resonators", *Phys. Lett. A*., **380**, pp. 525–529 (2016).

33. Asghari, M., Kahrobaiyan, M.H. and Ahmadian, M.T. "A nonlinear Timoshenko beam formulation based on the modified couple stress theory", *Int. J. Eng. Sci.*, **8**, pp. 1749-1761 (2010).

34. Lewandowski, R. and Wielentejczyk, P. "Nonlinear vibration of viscoelastic beams described using fractional order derivatives", *J. Sounds. Vib.*, **11**, pp. 228-243 (2017).

35. Wielentejczyk, P. Lewandowski, R. "Geometrically nonlinear, steady state vibration of viscoelastic beams", *Int. J. Non. Linear. Mech.*, **7**, pp. 177-186 (2017).

36. Roozbahani, M.M., Heydarzadeh Arani, N., Moghimi Zand, M. and Mousavi Mashhadi, M. "Analytical solutions to nonlinear oscillations of micro/nano beams using higher-order beam theory", *Sci. Iran. Trans. B Mech.*, **23** (5), pp. 2179-2193 (2016).

37. Alipour, A., Zand, M.M. and Daneshpajooh, H. "Analytical solution to nonlinear behavior of electrostatically actuated nano-beams incorporating van derWaals and Casimir forces", *Sci. Iran. Trans. B Mech.*, **22** (3), pp. 1322-1329 (2015).

38. Stojancovic, V. "Geometrically nonlinear vibrations of beams supported by a nonlinear elastic foundation with variable discontinuity", *Commun. Non. Linear. Sci. Num. Simul.*, **28**, pp. 66–80 (2015).

39. Szilard, R., *Theories and Applications of Plate Analysis: Classical, Numerical and Engineering Methods*, John Wiley & Sons, Inc. (2004).

40. Amabili, M., *Nonlinear vibrations and stability of shells and plates,* Cambridge University Press, New York, USA. (2008).

41. Meirovitch, L. *Fundamentals of Vibrations*, McGraw-Hill (2001)

42. Nayfeh, A.H. and Mook, D.T., *Nonlinear Oscillation*, John Wiley & Sons, Inc. (1995)

43. Amini, M.H., Soleimani, M., Altafi, A. and Rastgoo, A. "Effects of Geometric Nonlinearity on Free and Forced Vibration Analysis of Moderately Thick Annular Functionally Graded Plate", *Mech. Adv. Mater. Struct.*, **20**, pp. 709–720 (2013).

44. Rao, S.S. *Vibration of continuous systems*, John Wiley & Sons, Inc. (2007).

45. Asadi Dalir, M. and Seifi, R. "Direct method for deriving equilibrium equations in solid continuous systems", *Eng. Solid. Mech.*, **2**, pp. 321–330 (2014).

**Biographies**

**Mohammad Asadi Dalir** is a researcher in mechanical engineering and he is interested in nonlinear vibrations, vibrations of continuous systems, theory of plates and continuum mechanics. He is also interested in basic concepts of new mechanics, special and general relativity, quantum mechanics and its philosophical interpretation.

Table 1. The comparison of nonlinear natural frequency in Eq. (27) with those obtained from Galerkin method.

| Mode (s) | Method | $\lambda\omega_r$ ($r = 2s$) | $\omega_s$ | $\omega$ |
|---|---|---|---|---|
| 1 | Exact | 1292.3822 | 306.51536 | 1328.2332 |
|   | Galerkin | 1300.4345 | 306.51536 | 1332.0695 |
| 2 | Exact | 5169.5290 | 1226.0614 | 5312.9330 |
|   | Galerkin | 5201.7382 | 1226.0614 | 5344.2780 |
| 3 | Exact | 11631.440 | 2758.6382 | 11954.099 |
|   | Galerkin | 11703.911 | 2758.6382 | 12024.625 |
| 4 | Exact | 20678.611 | 4904.2458 | 21251.112 |
|   | Galerkin | 20806.953 | 4904.2458 | 21377.112 |
| 5 | Exact | 32309.356 | 7662.8841 | 33205.831 |
|   | Galerkin | 32510.864 | 7662.8841 | 33401.737 |

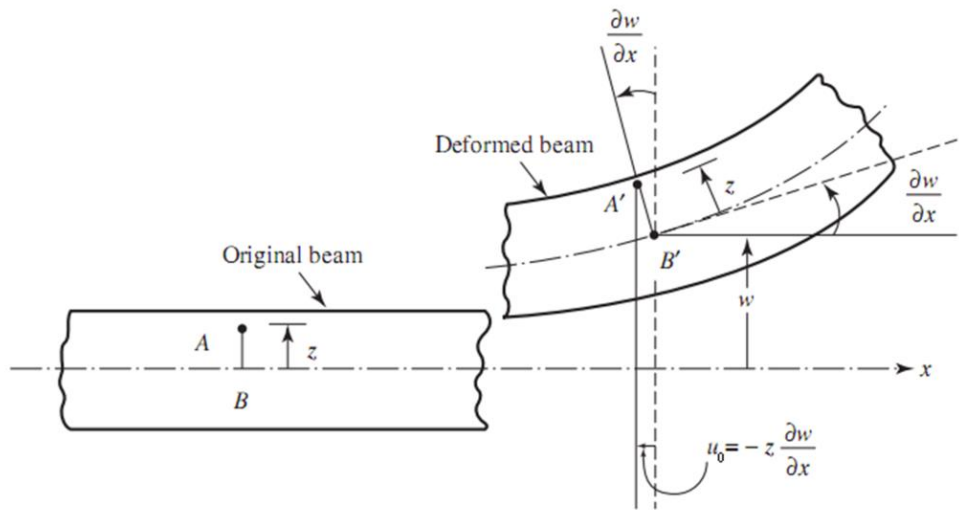

Figure 1. Geometry of a beam before and after deformation [44].

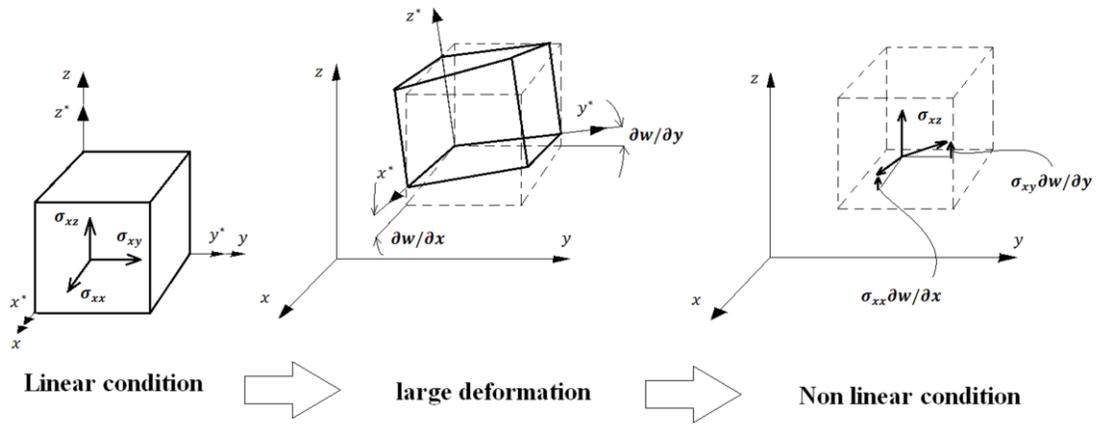

Figure 2. Effect of stress components for linear and nonlinear conditions in governing equations of three dimensional theory of elasticity [45]

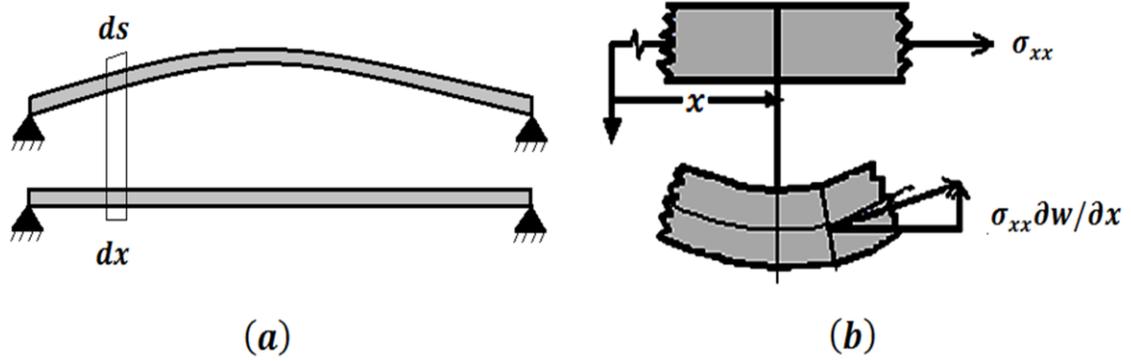

Figure 3. Stretch of beam develops in-plane stress which appears in transverse vibration.

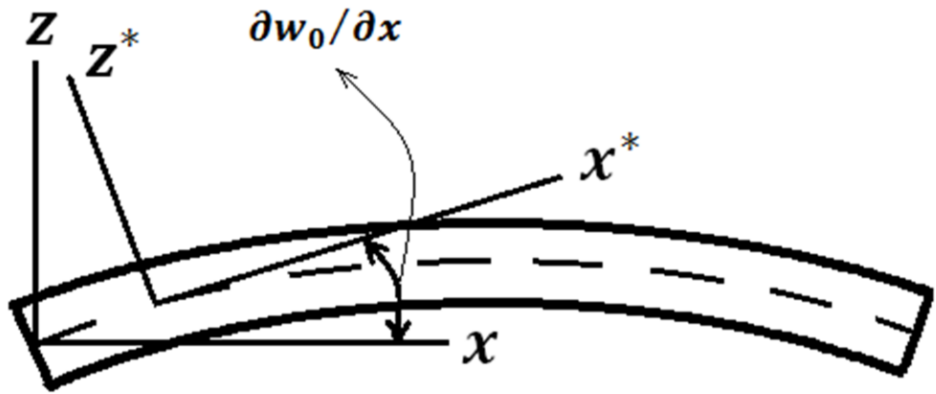

Figure 4. Rotation of Euler configuration with respect to Lagrange coordinates is equal to beam slope curve

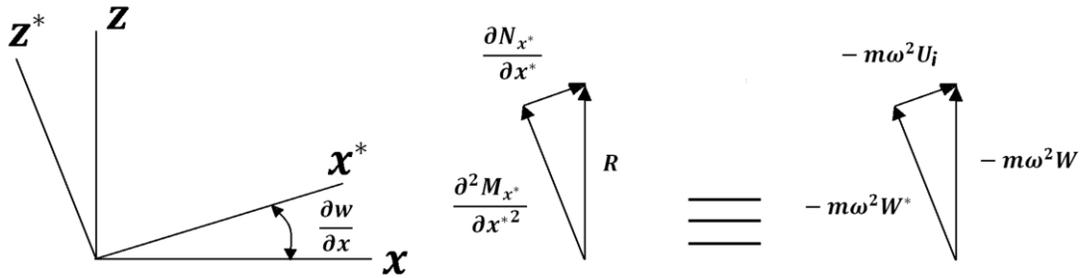

Figure 5. The nonlinear governing equation in Lagrange coordinates can be decomposed into two linear coupled equations in Euler coordinates

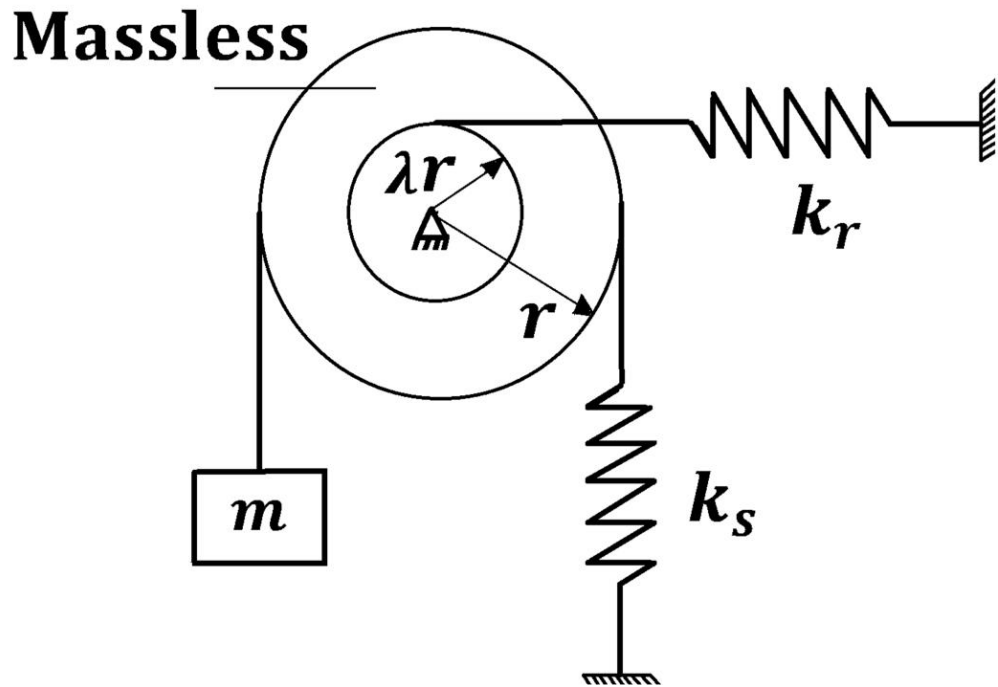

Figure 6. A schematic model of beam showing that its nonlinear stiffness is equal to linear combination of two linear springs

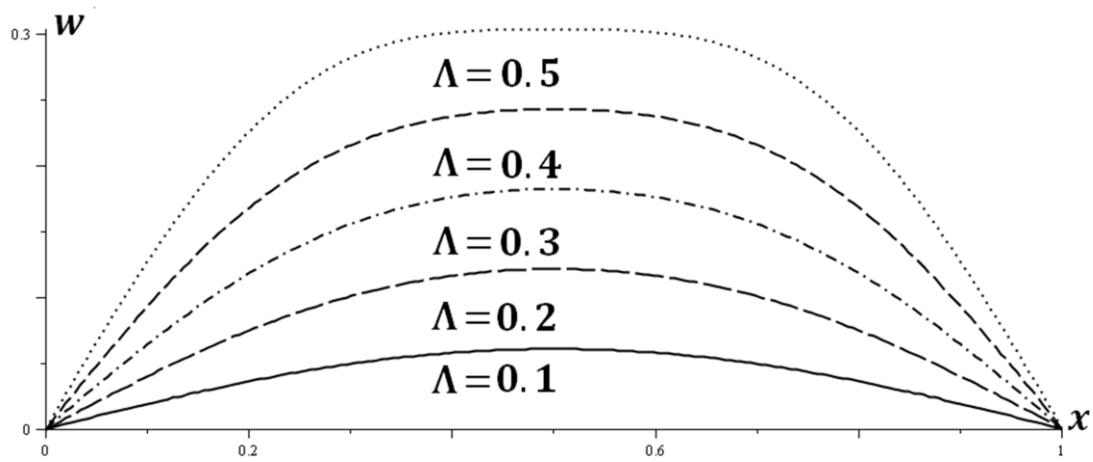

Figure 7. Nonlinear mode shapes tends to that of linear vibration when the amplitude is small

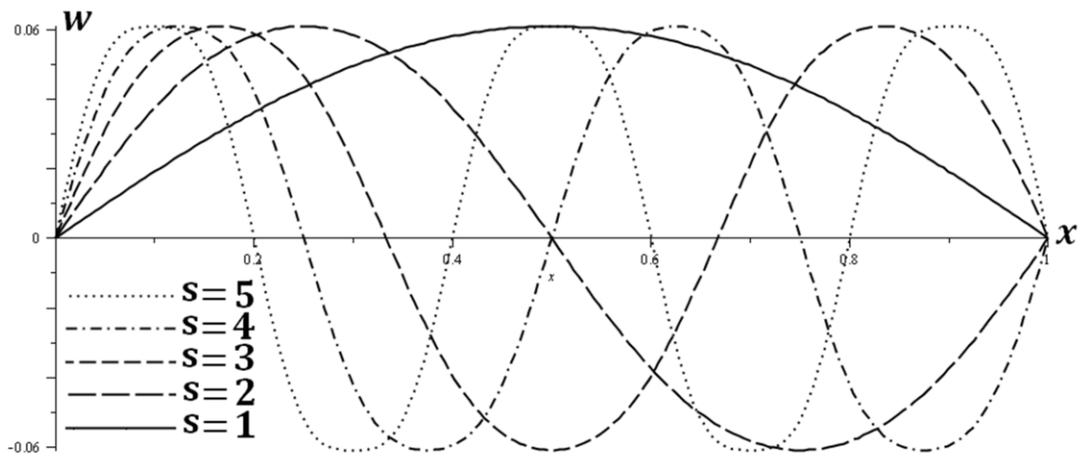

Figure 8. The nonlinear terms become more effective on mode shapes in higher modes of vibration

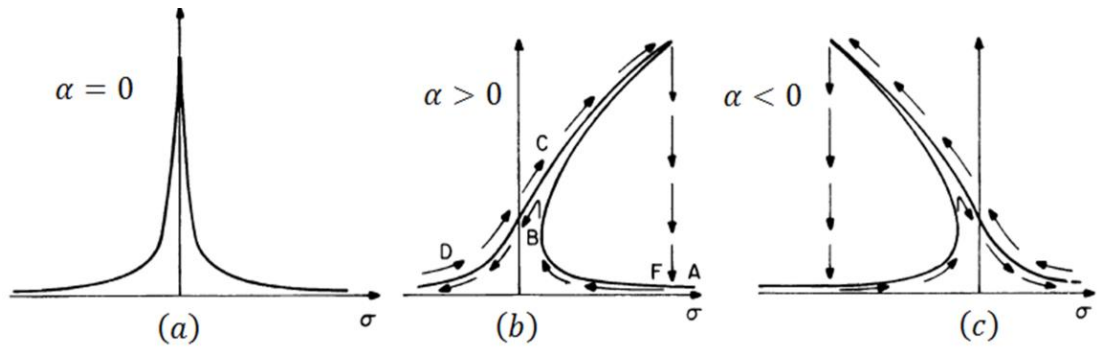

Figure 9. Frequency response of ($a$) linear, ($b$) hardening nonlinear and ($c$) softening nonlinear systems [31]

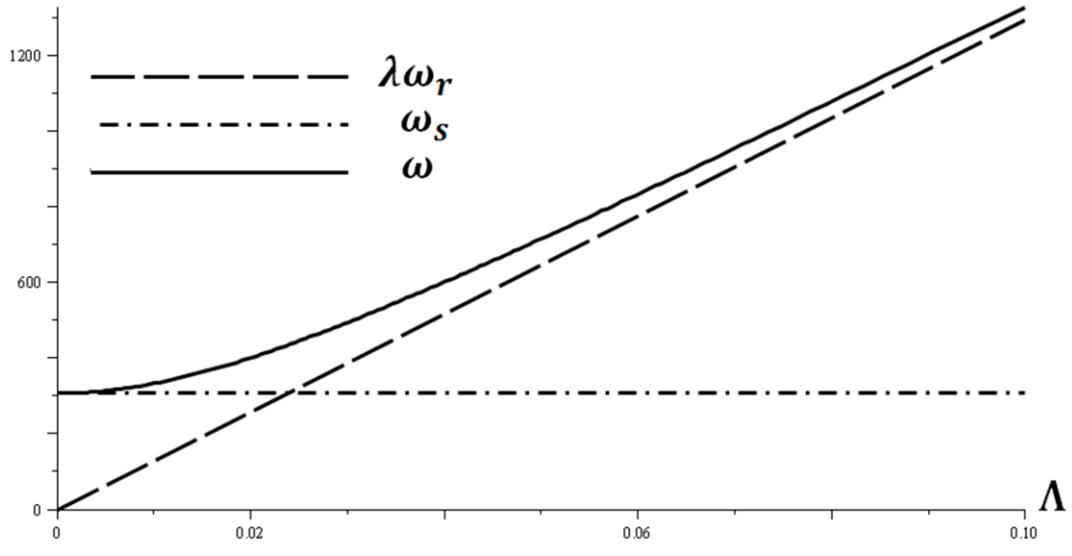

Figure 10. The effect of amplitude on nonlinear frequency

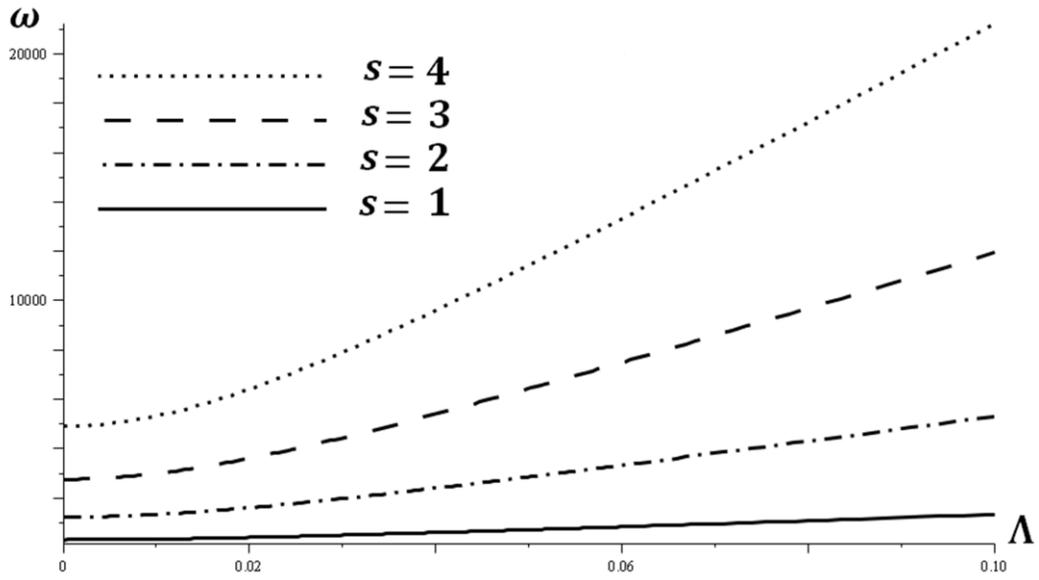

Figure 11. The effect of nonlinear terms on natural frequency increases, as the number of mode enhances

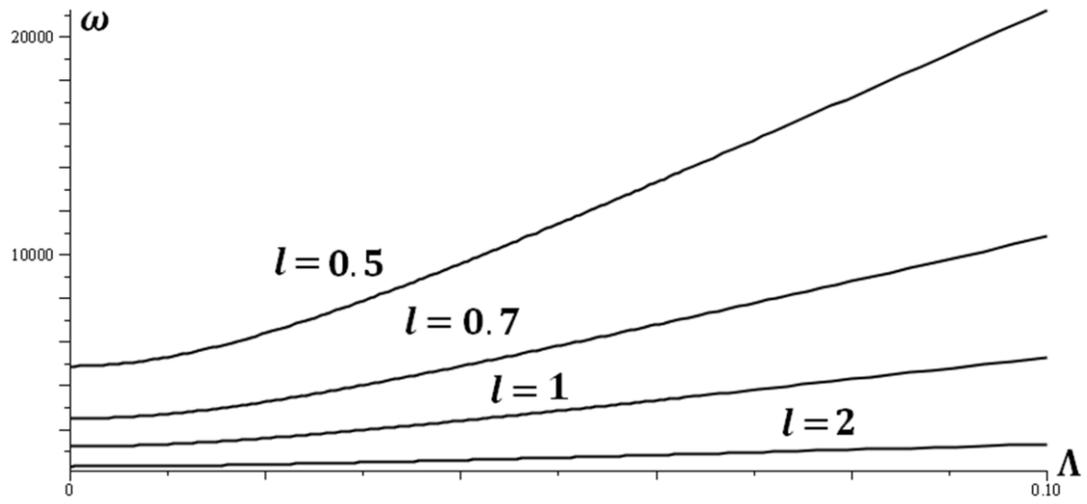

Figure 12. The effect of amplitude to length ratio on the nonlinear natural frequency

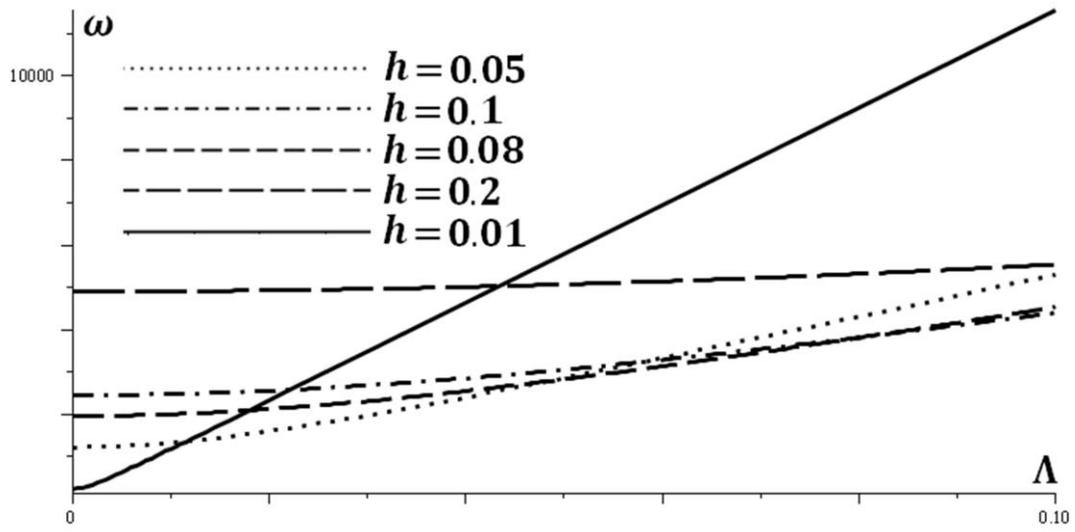

Figure 13. The effect of thickness on nonlinear natural frequency